# Integrated System for Malicious Node Discovery and Self-destruction in Wireless Sensor Networks

Madalin Plastoi

Politehnica University of Timisoara
Timisoara, Romania
madalin.plastoi@aut.upt.ro

Ovidiu Banias

Politehnica University of Timisoara
Timisoara, Romania
ovidiu.banias@aut.upt.ro

Daniel-Ioan Curiac

Politehnica University of Timisoara
Timisoara, Romania
daniel.curiac@aut.upt.ro

Constantin Volosencu

Politehnica University of Timisoara
Timisoara, Romania
constantin.volosencu@aut.upt.ro

Roxana Tudoroiu

Politehnica University of Timisoara
Timisoara, Romania
tudoroiu.roxana@ac.upt.ro

Alexa Doboli

State University of New York
Stony Broke, New York, USA
adoboli@ece.sunysb.edu

*Abstract* — **With the tremendous advances of the wireless devices technology, securing wireless sensor networks became more and more a vital but also a challenging task. In this paper we propose an integrated strategy that is meant to discover malicious nodes within a sensor network and to expel them from the network using a node self-destruction procedure. Basically, we will compare every sensor reading with its estimated values provided by two predictors: an autoregressive predictor [1] that uses past values provided by the sensor under investigation and a neural predictor that uses past values provided by adjacent nodes. In case the absolute difference between the measured and the estimated values are greater then a chosen threshold, the sensor node becomes suspicious and a decision block is activated. If this block decides that the node is malicious, a self-destruction procedure will be started.**

I. INTRODUCTION

With the continuous progress in micro-electro-mechanical systems (MEMS) and radio technologies, a new concept arose - wireless sensor networks (WSN). A wireless sensor network, being a collection of tiny sensor nodes with limited resources (limited coverage, low power, smaller memory size and low bandwidth), proves to be a viable solution to many challenging civil and military applications. Their deployment, sometimes in hostile environments, can be dangerously perturbed by any type of sensor failure or, more harmful, by malicious attacks from an opponent.

Sensor networks because of their specific limitations are susceptible to various kinds of attacks that cannot be prevented only by traditional methods (e.g. cryptography): eavesdropping, traffic analysis, selective forwarding, spoofing, wormhole attack, sinkhole attack, Sybil attack and Hello flood attack are the most significant [2]. But, almost certainly the most important danger, due to the inherent unattended characteristic of wireless sensor networks, is represented by node-capturing attack [3], where an enemy acquires full control over sensor nodes through direct physical contact. A node capturing attack is very feasible because of at least two reasons: a) practically, we cannot demand an efficient access control to thousands of nodes distributed in a large territory; and b) it is very difficult to assure tamper-resistance requirements because sensor nodes frequently need to be inexpensive to justify their use.

After an attacker gains the physical control over a sensor node he can extract secret information such as cryptographic keys to achieve unrestricted entrance to higher network levels, or by using reverse engineering techniques he can find security holes to compromise the entire sensor network.

Our proposed countermeasure is based on the fact that a corrupted node is better to be expelled from the network as soon as its malicious activity is started [4]. Even if more sensors are expelled, the WSN will function as designed because of one inherent feature: spatial redundancy [5].

In order to identify a corrupted sensor node, we presumed that even if it may still send authenticated messages (e.g., it can use the cryptographic keys already stored in its memory), it might not operate according to its original specifications sending incorrect readings to the base station. We will identify these sensors by using prediction techniques and will eliminate them by starting a self-destruction node procedure.

The rest of the paper is organized as follows. The second section presents a detailed description of our strategy. Section 3 presents the technique used for the self-destruction procedure applied to corrupted nodes. In the last two sections, experimental results and conclusions are offered.

## II. PREDICTION BASED METHODOLOGY

In a large number of applications where wireless sensor networks interact with sensitive information or function in hostile unattended environments, it is crucial to develop security related mechanisms.

Due to their nature, these networks have to resist to a plethora of possible attacks. The attackers will try to obtain in-network information or to corrupt the network partially or totally. For making this possible, the attackers will try to gain control over one or several network nodes. Our proposed defending strategy is based on the detection of malicious sensor nodes using predictors and the elimination of their effects by expelling them from the network using a self-destruction node technique.

### A. The Sensor Network Assumptions

In order to assure a high rank of efficiency for our malicious node detection and self-destruction strategy we chose a sensor network having the following features:

a) The sensor network is static, i.e., sensor nodes are not mobile; Moreover, each sensor node knows its own position in the field.

b) The base station, sometimes named access point, acting as a controller and as a key server, is supposed to be a laptop class device and supplied with long-term power. We also assume that the base station will not be compromised.

c) Between the three most common wireless topologies (star, mesh and cluster-tree) we chose a star topology (e.g. Cellular Wireless Network [6] and SENMA [7]) for our sensor network. Star topology is a point-to-point architecture where each sensor node communicates directly with the base station. The main characteristics of the star topology are: there are no node-to-node connections and no multi-hop data transmission; sensor synchronism is unnecessary; sensor do not listen, only transmit and only when polled for; complex protocols are avoided; dependability of individual sensors is much less significant. Because of these features, attacks on routing protocols (spoofing, selective forwarding, sinkhole attack, wormhole attack, Sybil attack and Hello flood attack) are almost impossible.

d) We rely on efficient secret-key cryptography with pre-distributed keys using Skipjack, RC5 or AES algorithms to encipher all data communications inside the sensor network; These three symmetric encryption algorithms have a common attribute that makes them an attractive alternative in case of sensor networks: they are able to encrypt short or medium size messages, like the ones send by sensors and received by base stations, in the case of limited power consumption. By using such appropriate cryptographic techniques the damaging potential of the passive attacks (eavesdropping and traffic analysis) can be ignored.

e) The measurements supplied by each sensor have an important deterministic character, rather than a strictly random (stochastic) one. In this case there exists a correlation between past values and the current one, giving us the power of prediction. As an example, the future values of temperature measurements in a location are strongly related with past and present values, so a prediction method can be applied.

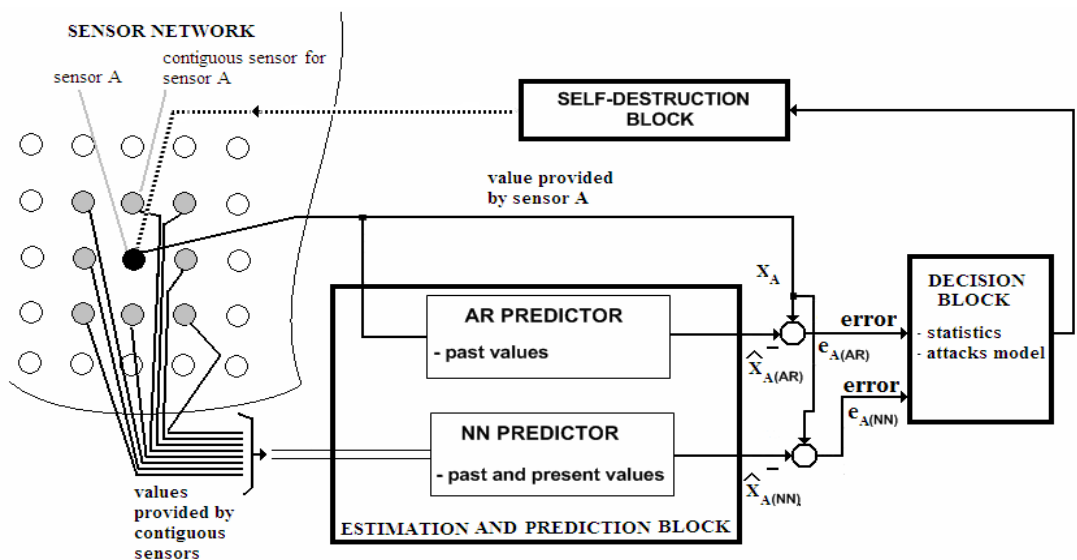

Figure 1. Malicious node discovery and self-destruction with mixed prediction

### B. The Proposed Strategy

Our strategy exploits two types of observations over the current and past measurements of the wireless sensor network nodes. First, the temporal redundancy in the sense that previous provided measurements from each sensor will be used to decide if the given sensor operates as desired or not. And second, based on the past and present values of neighboring nodes, the operation of the given sensor is classified as right or wrong.

We decided to use a mixed estimation and prediction system with two predictors – an autoregressive predictor (AR) and a neural network predictor (NN). The first predictor will describe the node evolution using its past values while the second one will describe node evolution using present and past values provided by neighboring nodes.

In Fig. 1 is presented the system architecture, composed of 4 major components: a) The Wireless Sensor Network under investigation; b) The Estimation and Prediction Block; c) The Decision Block; and d) The Self-destruction Procedure. The Estimation and Prediction Block includes two on-line predictors, one autoregressive predictor and one neural network predictor. The outputs of these two predictors represent the inputs for the Decision Block. At this level the system will provide automatic decisions upon engaging or not the self-destruction procedure for the given node.

In the following paragraphs we will present both predictors, including the mixed strategy.

Our stratagem to identify a corrupted sensor node is based on the fact that even if it may still send authenticated data it may not operate according to initial requirements, sending incorrect readings to the base station. These nodes will be identified in the moment they begin to broadcast erroneous information. For this purpose, we will present further two types of predictors, one autoregressive predictor, and one neural network predictor. Based on these predictors we will decide if a self-destruction procedure is needed.

In order to implement the first predictor, we considered that an autoregressive (AR) model efficiently approximates the evolution in time of the measurements provided by each sensor. An autoregressive or AR model describes the evolution of a variable only using its past values. This class of systems evolves due to its "memory", generating internal dynamics, and is defined as follows:

$$x(t) = a_1 \cdot x(t-1) + ... + a_n \cdot x(t-n) + \xi(t), \quad (1)$$

where $x(t)$ is the measurement series under investigation, $a_i$ are the autoregression coefficients, $n$ is the order of the autoregression and $\xi$ is assumed to be the Gaussian white noise. By convention, the time series $x(t)$ is assumed to be zero mean. If not, another term $a_0$ is added in the right member of equation (4). Establishing the correct model of order $n$ is not a simple task and is influenced by the type of data measurements and by computing limitations of the base station. Reasonable values of the order $n$ are between 3 and 6.

If the $a_i$ coefficients are time-varying, the equation (1) can be rewritten as:

$$x(t) = a_1(t) \cdot x(t-1) + ... + a_n(t) \cdot x(t-n) + \xi(t). \quad (2)$$

The model (2) can be used either to estimate the coefficients $a_i(t)$ in case the time series $x(t),...,x(t-n)$ are known (recursive parameter estimation), either to predict future value in case that $a_i(t)$ coefficients and past values $x(t-1),...,x(t-n)$ are known (AR prediction).

In parallel with the AR predictor, we use a three order feed forward neural network predictor with two hidden layers. The input layer consists of neurons which are associated with values provided by adjacent nodes at moments *t*, *t-1* and *t-2*. The output layer has only one neuron for the estimated value.

For network's node A, the estimated value by the neural predictor is:

$$\hat{x}_{A(NN)}(t) = f(X_{A,adj}(t-1),...,X_{A,adj}(t-n), B_{A(NN)}), \quad (3)$$

$$X_{A,adj}(t-i) = (x_{A,adj1}(t-i),...,x_{A,adjm}(t-i))^T, \quad (4)$$

$$B_{A(NN)} = (b_{A,adj1},...,b_{A,adjm})^T, \quad (5)$$

where: $x_{A,adjk}(t-i)$ is the value from the adjacent node k at $t-i$ moment, $B_{A(NN)}$ is a vector containing the trust factors of each of the *m* adjacent nodes of A, and *n* is the predictor order.

The trust factor $b_{A(NN)}$ has a linear dependence on the previous node trust factor and the error value. However, the computation of $b_{A(NN)}$ is made in the decisional block.

After the node's output of the neural prediction, an error value is obtained:

$$e_{A(NN)}(t) = |x_A(t) - \hat{x}_{A(NN)}(t)|. \quad (6)$$

*C. The Autoregresive Prediction*

Our strategy exploits the temporal redundancy in the sense that previous provided measurements from each sensor will be used to decide if the sensor operates as desired or not. The plan is the following: an attacked sensor node that will attempt to insert false information into the sensor network will be recognized by comparing its output value with the value predicted using past readings offered by that specific sensor (Fig. 2). In the case that any malicious activity is observed, the Decision Block is triggered to decide if the self-destruction procedure must be started for this specific node in order to prevent its further undesired activity.

The complete mechanism workflow runs while network is active. By considering a specific node symbolized by A, this process is done in the following steps:

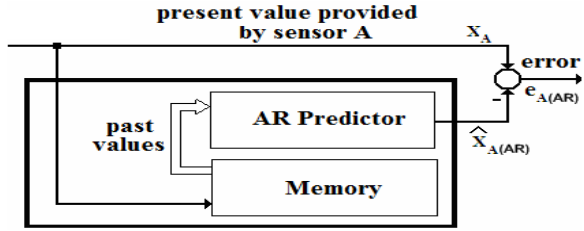

Figure 2. Autoregressive predictor

a) Associate a node trust indicator with every sensor node. The specified sensor node A will have a trust factor denoted by $b_{A(AR)}$. This integer value is initially set to zero ($b_{A(AR)} = 0$ for a fully reliable sensor node) and is incremented during our methodology every time a potential malicious activity is encountered. This trust factor must be reset to zero, if no potential malicious activity is encountered for a long period of time. The node trust indicator represents, in other words, the perception of confidence between the WSN and that specific sensor.

b) At every moment of time $t$, estimate the present value $\hat{x}_{A(AR)}(t)$ provided by sensor node A, using the past readings $x_{A(AR)}(t-i)$ provided by the same sensor A. For the sensor A, we can write:

$$\hat{x}_{A(AR)}(t) = f(x_{A(AR)}(t-1),...,x_{A(AR)}(t-n)) \quad (7)$$

where $n$ is the estimator's order. In our approach, an on-line AR predictor performs this step.

c) Compare the present value $x_A(t)$ measured by the sensor node A with its estimated value $\hat{x}_{A(AR)}(t)$ by calculating the error:

$$e_{A(AR)}(t) = x_A(t) - \hat{x}_{A(AR)}(t) \quad (8)$$

d) Increment the node trust indicator $b_A$ if the error $e_{A(AR)}(t)$ exceeds a given threshold $\varepsilon_{AR}$: $|e_{A(AR)}(t)| > \varepsilon_{AR}$ (this is done only one time inside a transitory time zone – due to the internal structure of recursive predictors, an error obtained at instant time $t$, is propagated/attenuated in the recursive predictor response for some instants, obtaining a transitory regime). If the node trust indicator is higher than a chosen value $\gamma$ ($b_{A(AR)} > \gamma$), the node could be declared as potentially malicious and depending on the Decision Block output, a self-destruction procedure could be started for the A node.

The associated pseudocode is presented in Fig. 3.

```
//this function is performed for each node in the network
int Autoregression(int nodeId)
{
SET b_{A(AR)}, ε_{A(AR)} ; //node trust indicator, threshold
WHILE (network is active)
{
...
x_A =READ sensor A; //get sensor actual value for node ID equal
                    //with nodeID
...
x̂_{A(AR)} = ARpredict(prior x_A values); //call AR prediction
e_{A(AR)} = x_A − x̂_{A(AR)} ; //calculate the error
IF (ABS ( e_{A(AR)} ) > ε_{A(AR)} )
{
  IF (AR predictor is not in transitory regime)
  { b_{A(AR)} = b_{A(AR)} + 1; //increment node trust indicator
    START thread TRANSITORY_REGIME;
      //a counter set on k //and will be decremented every
      //instant until it becomes zero
  }
  DECISION_BLOCK (node with node ID equal to nodeID);
    //call //decision method
}}
...
}
```

Figure 3. AR implementation pseudocode

First of all, we have to associate a threshold $\varepsilon_{AR} > 0$ for every sensor node. This threshold will be used to decide if a sensor operates normal or abnormal and its measured value depends on the type of the sensor and its specific and desired operation in real environment. For a specific sensor A, the threshold will be denoted by $\varepsilon_{A(AR)}$.

After this initialization, at every instant $t$, we will compute the estimated value $\hat{x}_{A(AR)}(t)$ relying only on past the values $x_A(t-1),...,x_A(0)$ and we will use both parameter estimation and prediction as presented in the following steps:

*First step*: we will estimate the parameters $a_i(t)$ using a recursive parameter estimation method. From a large number of methods for estimating AR coefficients we decided to use a numerically robust RLS (recursive least square) variant based on orthogonal triangularization, known in literature as QRD-RLS [8]. One of the reasons is that it can be implemented efficiently on the base stations level (laptop class device).

*Second step*: we will obtain the prediction value $\hat{x}_{A(AR)}(t)$ using the following equation:

$$\hat{x}_{A(AR)}(t) = a_1(t) \cdot x_A(t-1) + ... + a_n(t) \cdot x_A(t-n) + \xi(t) \quad (9)$$

245

The corresponding pseudocode for implementing the estimation procedure is presented in Fig. 4:

```
float ARpredict(prior x_A values)
{ CALCULATE autoregression coefficients a_i ;
    // an estimation using QRD-RLS method
  CALCULATE predicted value  x̂_{A(AR)} ;
  //compute sensor predicted value as a result of (9)
  RETURN  x̂_{A(AR)} ; }
```

Figure 4. Autoregressive prediction pseudocode

Following, we will compare the present value $x_A(t)$ measured by the sensor node with its estimated value $\hat{x}_{A(AR)}(t)$, and the error $e_{A(AR)}$ will be computed using equation (8).

### D. The neural network prediction

In order to improve the efficiency and reliability of the Estimation and Prediction Block, we implemented a neural network predictor based on measurements provided by contiguous sensors [9]. This predictor will work in parallel with the autoregressive predictor to discover all malicious activity. In Fig. 5 the neural network predictor is presented:

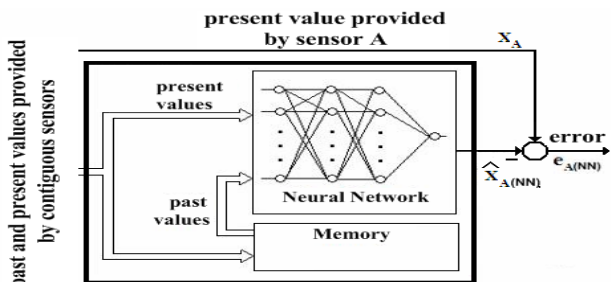

Figure 5. Neural Network predictor

A neural network predictor consists of three or more layers of artificial neurons (Fig. 6). Usually, neural networks have at least three layers, an input layer, an output layer and a hidden one. Our implemented neural network has four layers because we used two hidden layers. Neurons are linked one with each other through quantitative relations known as weights. Each layer is activated by an activation function. Most used activation functions are: direct identity (for input layers), sigmoid, hyperbolic tangent and linear functions. In fact this neural predictor model exposes a composition function which describes how the inputs are transformed into outputs.

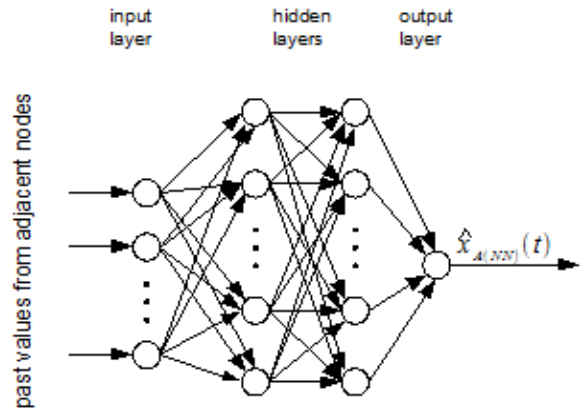

Figure 6. Neural network predictor

Each neuron implements the following function:

$$f(x) = K(\sum_i \omega_i g_i(x)), \qquad (10)$$

where K is a composition function, $\omega_i$ are the weights, and $g_i$ is a vector containing neuron's inputs $g = (g_1, g_2, ..., g_n)$ [10].

In order to use a neural network predictor, two steps must be performed:

*Neural network training* - will be established the number of hidden neurons and the neural network weights $\omega_i$ by performing successive training sessions using Levenberg-Marquardt method [11]. For the hidden layers we will use hyperbolic tangent activation function and for the output layer we will use a linear activation function. This training step is done off-line, prior to the neural network predictor implementation on the base station.

*Neural network on-line prediction* - will be obtained the prediction value $\hat{x}_{A(NN)}(t)$, computing for each neuron the equation (10). This procedure starts with the neurons from the input layer and ends with the neuron from the output layer and it's implemented as an on-line predictor on the base station. In the end, the error $e_{A(NN)}$ is computed using equation (6). The associated pseudocode for the neural network predictor is presented in Fig. 7:

```
//this function is performed for each node in the network
int NNPrediction(int nodeId)
{
SET b_{A(NN)}, ε_{A(NN)} ; //node trust indicator, threshold
WHILE (network is active)
{
 …
 x_A =READ sensor A; //get sensor actual value for node ID equal
 //with nodeID
 …
```

246

```
x̂_A(NN) = NNpredict (present and past adjacent nodes values);
    //call NN prediction
e_A(NN) = x_A − x̂_A(NN) ;//calculate NN error
IF (ABS ( e_A(NN) ) > ε_A(NN) )
  IF (NN predictor is not in transitory regime)
  { b_A(NN) = b_A(NN) + 1  //increment node trust indicator
    START thread TRANSITORY_REGIME;
        //a counter set on k //and will be decremented every
        //instant until it becomes zero
  }
  DECISION_BLOCK (node with node ID equal to nodeID);
    //call decision method; Fig. 8.
}
…
}
```
Figure 7. NN implementation pseudocode

The initialization steps are the same as for the AR prediction, a threshold $\varepsilon_{A(NN)} > 0$ being associated with sensor node A.

At every instant *t*, we will compute the estimated value $\hat{x}_{A(NN)}(t)$ relying on the present and past values of the adjacent neighbors of the node A.

### E. The Decision Block

The Decision Block will provide the decision to start the self-destruction procedure of a malicious node, based on the pair of errors, ($e_{A(AR)}(t), e_{A(NN)}(t)$) that will be used as inputs in an expert system, by computing two trust indicators $b_{A(AR)}$ and $b_{A(NN)}$.

These trust indicators are initially set to zero ($b_{A(AR)} = 0, b_{A(NN)} = 0$) signifying that in the beginning, the sensor is considered to be fully reliable. After that, when potential malicious activity is detected, the trust indicators are incremented. In order to filter possible sporadic malfunctions of the sensors, these counters must be reset to zero at specific intervals in time.

An efficient Decision Block can be implemented either as a fuzzy-based system, either as a rule-based system. In this paper we have chosen a rule-based approach. The Decision Block architecture is presented in the Fig. 8:

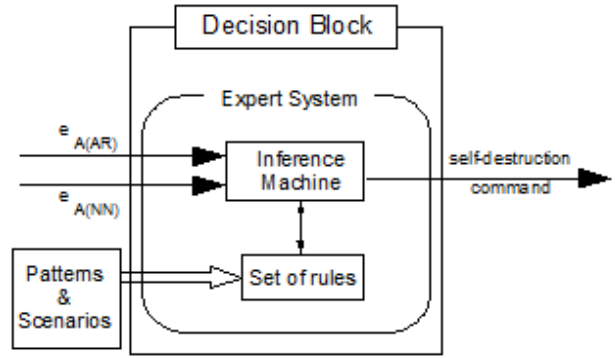

Figure 8. Decision Block

Based on the previously defined patterns & scenarios, the decision block was implemented as an expert system. We defined two parameters $\alpha, \beta$, with $0 < \alpha < \beta$, for classifying each of the trust factors $b_{A(AR)}, b_{A(NN)} \in \mathbb{N}$, as follows:

a) *5 categories for* $b_{A(AR)}$

$$b_{A(AR)} = 0; b_{A(AR)} \in (0,\alpha); b_{A(AR)} = \alpha; \qquad (11)$$
$$b_{A(AR)} \in (\alpha,\beta); b_{A(AR)} = \beta$$

b) *5 categories for* $b_{A(NN)}$

$$b_{A(NN)} = 0; b_{A(NN)} \in (0,\alpha); b_{A(NN)} = \alpha; \qquad (12)$$
$$b_{A(NN)} \in (\alpha,\beta); b_{A(NN)} = \beta$$

Using this classification, we developed 25 types of rules for the pair ($b_{A(AR)}, b_{A(NN)}$) having the inputs $b_{A(AR)}$, $b_{A(NN)}$, $\alpha$, $\beta$, and the outputs the activation or not of self-destruction procedure. The rules formalization is presented in Fig. 9:

RULE X: // activation of self-destruction procedure

if Evaluate($b_{A(AR)}, b_{A(NN)}, \alpha, \beta$) then
    Self_Destruction( sensor A);

RULE Y: // not activation of self-destruction procedure

if Evaluate($b_{A(AR)}, b_{A(NN)}, \alpha, \beta$) then
    DoNothing();

Figure 9. Rule definition

As presented in Fig. 9, the activation of certain rules lead to the self-destruction procedure of the given node (Self_Destruction procedure). This procedure is presented in the following paragraph.

## III. MALICIOUS NODE SELF-DESTRUCTION

If a certain node has been tagged as malicious, the base station will initiate a self-destruction sequence for that specific node. The self-destruction routine is divided into several actions:
- Erase node RAM memory that contains susceptible network information, driven software and cryptographic keys and also other additional memories (e.g. flash memory, if present);
- Drain node battery in different ways like R/T radio flood or node logical unit infinite cycle [12];
- Destroy node radio device;
- Delete node unique identifier from the lists of each of the neighbor nodes, including base station; This way an already captured node won't gain authentication rights if an attacker tries to reintroduce it in the network (will disable auto-organization property);
- Mask node measurement nature by hiding the type of the sensor that has been used (each node has one or more sensors and knows in a logical way which of them is used for measurements).

The above actions have to be performed in order of their importance, although some kind of concurrency could be assured. For example: the initiation of self-destruction could start the procedure for draining node battery, but in the same time it could conduct erasing actions for memory and cryptographic keys.

Self-destruction should take into consideration all network characteristics from design to deployment including the topology. Also it strongly depends on the node hardware profile. For the proposed star network model, self-destruction will imply only the base station and the compromised node – as we stated earlier, each node communicates directly with the base station.

Basically, self-destructive sequence may be a software routine embedded into node's memory or sent bit by bit from the base station to the aimed node. Entire code has to be compatible with nodes and base station operating system. The pseudocode for node self-destruction is presented in Fig. 10:

```
void Self_Destruction (sensor A)
WHILE (sensor A is in network)
{
START thread
 CONSUME battery energy //broadcast specific messages;
START thread
 {ERASE node memory; //erase RAM and flash memory
  DISABLE auto-organization property;
          //delete node identifier from all neighbor's lists
  DESTROY node radio device;
  MASK node measurement nature;
          //for hiding the type of the sensor
}}
```

Figure 10. Node self-destruction pseudocode

In the optimistic scenario, after self-destructive routine was initiated, the intended node is destroyed and obvious, undetectable in the network. All references to its identifier will disappear from all network devices. Cryptographic keys and stored information will be also deleted.

In the pessimistic scenario, the self-destructive response won't have any triggered action attached to it; the corrupted node will still be alive in a fully or partially functional state. This scenario will have to be avoided by reducing the probability of some unfortunately events like:
1. Self-destructive routine was not suitable implemented for node hardware profile;
2. The node's software is modified by the enemy in such way that it doesn't accept incoming messages from the base station, it only sends malicious data;
3. Node battery is almost exhausted at the moment of executing the routine. In this case no memory erasing will be performed. The attacker will replace the batteries and the node will be partially or totally running.

Our solutions for avoiding these incidents are: testing of the self-destruction code on every type of sensor nodes that is included in the WSN; hiding the self-destruction routine into the node's memory; paying attention to the energy consumption on each sensor.

## IV. CASE STUDY

For validating our strategy, we used a Crossbow wireless sensor networks, containing 15 Mica2 nodes and one base station in a star topology. The sensors were set to measure the temperature in a room. In order to model an attack upon sensor A, we made the following experiments:
- First case study: at specific moments in time we intervened with an electric incandescent lamp (200 watts) placed very near to sensor A, for disrupting the normal functioning of only one specific node;
- Second case study: at specific moments in time we placed a heat source near to the network. In this case, normal functioning was disrupted for several network nodes, including the sensor node A.

We set up both predictors with the order $n = 3$, the thresholds $\varepsilon_{A(AR)} = 1°C$, $\varepsilon_{A(NN)} = 2°C$ and the pair $(\alpha, \beta) = (3, 5)$.

We considered that the node A has 8 adjacent neighbors. Past and present values from these adjacent neighbors will be used for neural network prediction of the node A value at a certain moment in time, while past values of node A will be used for autoregressive prediction.

The predictor contains 4 layers: one for input with 24 neurons, two hidden layers and one output layer for the estimated value. The hidden layers have 48 and 24 neurons.

The neural network predictor was trained using three arrays of inputs corresponding to the 8 adjacent neighbors at three moments of time (predictor order) and a target value for node A (Fig. 11). The training was performed offline, using the Matlab toolbox functions and features (e.g. train function, nntool UI) [13].

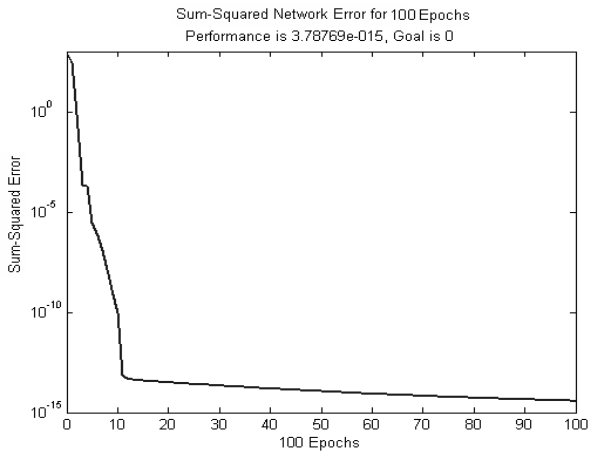

Figure 11. Neural network training session

After the training step was finished, the neural network predictor was installed on the base station node.

*A. First case study*

In this case study, we simulated an attack over a certain node, by inserting a heated lamp in its close neighborhood, leaving the other nodes unaffected (Fig. 12). We observed the result of both predictors and the system's decision making.

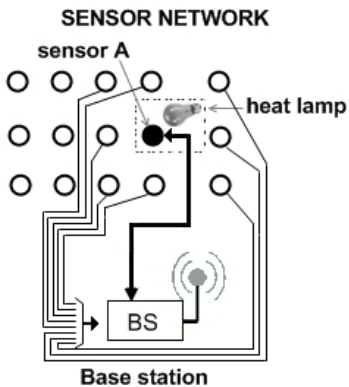

Figure 12. First case study description

In Fig. 13a we presented the sensor's A output time series, including our three "malicious" interventions at instants t=15sec, t=20 sec and t=27 sec. In Fig. 13b and 13c we presented the time variation of the AR and NN predicted time series, and in Fig. 13d and 13e we presented the evolution in time of the errors $e_{A(AR)}(t)$ and $e_{A(NN)}(t)$.

As node A is the only attacked node, the NN predictor estimated value remains in the limits of the adjacent nodes real values, while the AR predictor estimated value is in the vicinity of the attacked node real values, $e_{A(NN)}(t) > e_{A(AR)}(t)$, $t \in \{15, 20, 27\}$. From the graphics depicted in the Fig. 13d and 13e, we can observe that the trust factors $b_{A(AR)}$ and $b_{A(NN)}$ have the same value, $b_{A(AR)} = b_{A(NN)} = 3$. The decision to engage the self-destruction procedure is made based on a predefined rule:

If ($b_{A(NN)} == \alpha$ AND $b_{A(AR)} == \alpha$)
  Self_Destruction(sensor A);

The results are as expected: after exceeding the thresholds $\varepsilon_{A(AR)}, \varepsilon_{A(NN)}$ for three times for both predictors (Fig. 13d, Fig. 13e), the sensor A is expelled from the WSN by starting its self-destruction procedure. This result can be observed in Figure 13 where no more readings are obtained from sensor A after t=27 sec.

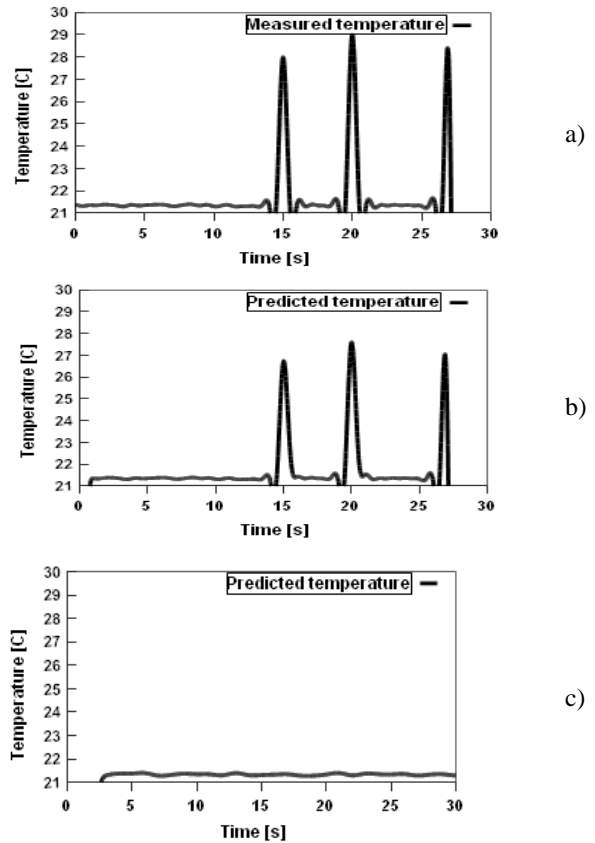

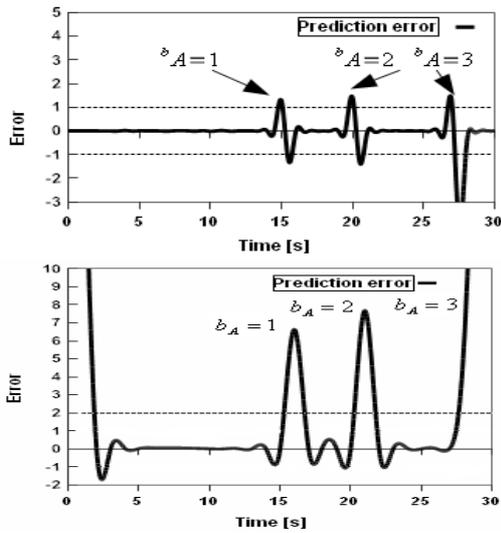

Figure 13. a)The sensor's output time series;
b)AR Predicted time series; c) NN Predicted time series;
d) AR Prediction error time series; and e) NN Prediction error time series;

### B. Second case study

In this case study, a heat source is activated at time moments t=15sec, t=20 sec, and t=27 sec in order to affect all sensor nodes, which represent a normal operating scenario - it is not a simulation of an attack. All network nodes are affected by a hot air wave.

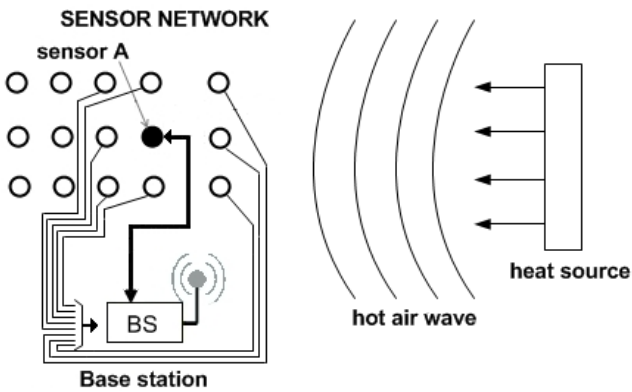

Figure 14. Second case study description

The evolution in time of the AR predicted values and the evolution of the error $e_{A(AR)}(t)$ are similar to the previous case study.

Assuming that node A has 8 adjacent neighbors, since all these neighbors are also affected by this heat wave (Fig. 14), the neural network predicted time series for the node A will have a similar evolution with the AR predictor time series (Fig. 15b, Fig. 15c). The computed error for the NN predictor case is lower than the computed error for the AR predictor case, $e_{A(NN)}(t) < e_{A(AR)}(t), \ t \in \{15, 20, 27\}$. Also the error $e_{A(NN)}(t)$ is lower than the threshold $\varepsilon_A = 2^\circ C$. The decision to engage or not the self-destruction procedure is made based on a predefined rule:

if ( $b_{A(NN)} == 0$ AND $b_{A(AR)} == \alpha$ )
  DoNothing();

By activating this rule, the Decision Block decided that no attack was performed over the node A.

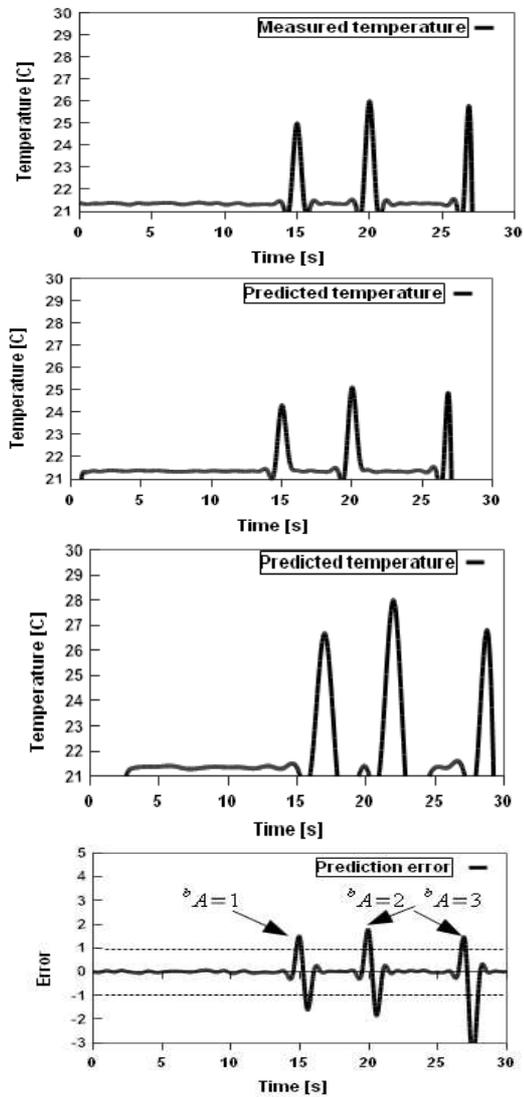

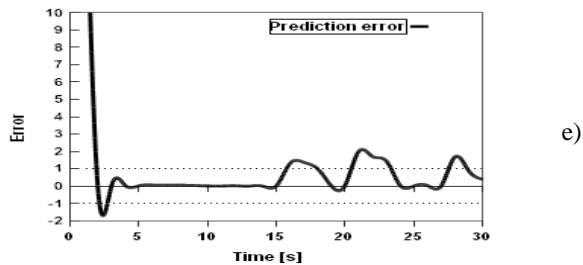

Figure 15. a)The sensor's output time series;
b)AR Predicted time series; c) NN Predicted time series;
d) AR Prediction error time series; and e) NN Prediction error time series;

Due to the diversity of the attack patterns we proved that the use of the two predictors in parallel is more accurate than the usage of a single predictor. For example, the use of only one predictor (AR predictor) in the second case study can lead to a wrong result – the self destruction of a normal-functioning node.

## V. CONCLUSIONS

Security issues related to WSN become more and more an important research area. Detecting abnormal/malicious operation of motes and offering efficient countermeasures represents a difficult task. In this paper we propose a combined strategy that not only detects the corrupted nodes, but also excludes their malicious activity using a self-destruction node technique. Due to the inherent spatial redundancy feature of WSN, applying a self-destruction procedure to the corrupted nodes has no major inconveniences, extending the secure operation of the entire sensor network.

The integrated system presented in this paper has some noticeable advantages: decision for node self-destruction is taken based on two predictions, two errors and two trust factors, therefore the accuracy is better; the whole system takes into consideration not only the evolution of a specific node, but also its neighbors evolution; as the two predictor models have a different nature, different attack patterns could be treated. Our strategy has also a drawback: a bigger computational power is needed at the base station level.

ACKNOWLEDGMENT

This work was developed in the frame of PNII-IDEI-PCE-ID923-2009 CNCSIS - UEFISCSU grant and was partially supported by the strategic grant POSDRU 6/1.5/S/13-2008 of the Ministry of Labor, Family and Social Protection, Romania, co-financed by the European Social Fund – Investing in People.